\documentclass{elsart}
\usepackage{graphicx}
\usepackage{amsfonts}
\usepackage{amssymb}
\usepackage{amsmath}
\usepackage[all]{xy}
\begin{document}
\begin{frontmatter}

\title{Non-commutative geometry inspired  higher-dimensional charged,
black holes\thanksref{infn0}}\thanks[infn0]{
As this paper was nearing completion our friend
Gallieno Denardo has suddenly passed away. As his former students, we dedicate
this paper to his memory and strongly believe that without his early influence
it probably  would have never been written.}
\author{Euro Spallucci\thanksref{infn}}
\thanks[infn]{e-mail address: spallucci@ts.infn.it }
\address{Dipartimento di Fisica Teorica, Universit\`a di Trieste
and INFN, Sezione di Trieste, Italy}

\author{Anais Smailagic\thanksref{infn2}}
\thanks[infn2]{e-mail address: anais@ts.infn.it }
\address{INFN, Sezione di Trieste, Italy}

\author{Piero Nicolini\thanksref{infn3}}
\thanks[infn3]{e-mail address: nicolini@cmfd.univ.trieste.it}
\address{Dipartimento di Matematica ed Informatica, Consorzio di 
Magnetofluidodinamica, Universit\`a di Trieste 
and INFN, Sezione di Trieste, Italy,
Department of Physics, California State University Fresno, Fresno, CA 93740-8031}

\begin{abstract}
We obtain a new, exact, solution of the Einstein's equation in higher 
dimensions.
The source is given by a static spherically symmetric, Gaussian distribution
of mass and charge. 
The resulting metric describes a regular, i.e. curvature singularity free,
 charged black hole in higher dimensions.  The metric smoothly
 interpolates between Reissner-Nordstr\"om geometry
at large distance,  and deSitter spacetime at short distance.
 Thermodynamical properties of the black hole are investigated and
 the form of the Area Law is determined. 
 We study pair creation and show that the upper bound
on the discharge time increases with the number of extra dimensions.
\end{abstract}
\end{frontmatter}

 \section{Introduction}
  By now, it can be safely claimed that string theory induced
  non-commutative geometry \cite{Witten:1995im}%,Seiberg:1999vs},
  provides an effective framework to study short distance
  spacetime dynamics. The original
  idea, revived by the today stringy formulation, dates back to
  Snyder seminal paper \cite{Snyder:1946qz}. In this model
  there exist a universal minimal length scale. At distances near
  $\sqrt\theta$ the classical concept of smooth
  spacetime manifold breaks down. It is generally assumed that  $\sqrt\theta$
  is closed to the Planck length, and as such it would be unaccessible both
  to present and future experimental observations.\\
  A promising alternative to the Planck scenario has been proposed, recently,
  based on  the presence of ``large
  extra-dimensions''. It allows to  lower unification scale, i.e. string
  tension, down to
  $TeV$ energy so that stringy effects could be soon observed
  at LHC \cite{ArkaniHamed:1998rs}%,Antoniadis:1998ig}.  
  Thus,if non commutative geometry is induced by string theory, the 
  corresponding length scale should be lowered as well.
  These ideas offer exciting , near future, possibility of experimentally
  probing both non-commutativity and quantum gravity effects.
   The most convincing confirmation of $TeV$ \textit{Quantum Gravity} would
   be the production of a mini black hole (BH)  in hadronic collisions
 \cite{Giddings:2001ih},%Bleicher:2001kh,Stenmark:2002yb,Hsu:2002bd},
%\cite{Cavaglia:2003qk,Chamblin:2004zg,Cavaglia:2004jw,Shankaranarayanan:2003qm},
%\cite{Rizzo:2006uz,Rizzo:2005jz},
\cite{Casanova:2005id},\cite{Alberghi:2006qr}%,Casadio:2001wh},
\cite{Stoecker:2006yz}. An unambiguous signature of such an event
   is based on a detailed analysis of both decay products and of the
   eventual BH remnant. So far the standard scenario
   of BH evaporation  offered a detailed and definite predictions
   for the spectrum of emitted particles, while it is  inconclusive
   about the final phase of BH evaporation.  This ambiguity can be traced back
   to the semi-classical description of Hawking process
   in the sense that matter emitted by BH is
   represented by  quantum field theory in curved spacetime, while BH
   itself is still a \textit{classical} background geometry. On the other hand,
   the final stage of BH decay requires quantum
   gravity corrections which the semi-classical model is unable to
   provide. \\
   In view of the ``quantum'' character of non-commutative
   geometry, one expects that in this model the late stage of the evaporation
   is determined by spacetime fluctuations when the radius of the BH horizon
   becomes comparable with $\sqrt\theta$. Usual attempts to obtain BH solutions 
   of  ``non-commutative'' Einstein's equations are based upon perturbative
   expansion in the $\sqrt\theta$ parameter.  So far, all these attempts lead
   to unconvincing results: it is unacceptable that curvature singularities
   can survive coordinate fluctuations in spite of the
   existence of a minimal length 
   \cite{Nasseri:2005ji}.%,Nasseri:2005yr,Kar:2005qe}
   %\cite{Chaichian:2007dr,Chaichian:2007we,Kobakhidze:2007jn,Mukherjee:2007fa}.
   It is likely that these problems are due to the method of
   approximating the original, intrinsically non-local theory, with a local
   truncated expansion of the action functional in terms of derivative
   interactions.\\
   Against this background, we proposed an approach based on
    ``quasi-classical coordinates'', keeping track
    of the intrinsic non-locality in the form a  minimal length in the
    spacetime fabric.
   As a result, matter/energy densities are described by
   minimal width Gaussian  distributions, in complete agreement with the basic
   principles of quantum mechanics. Solutions of Einstein's equations with
   such  smeared sources give new kind of regular BHs
  in four \cite{Nicolini:2005vd}%Ansoldi:2006vg}
  and higher  dimensions \cite{Rizzo:2006zb}.
  These models  lead to new predictions for the final stage of
  Hawking evaporation.  End result is a massive remnant in the form of
   a neutral, cold, near-extremal BH. The mass of this residual and
   stable object is determined by the $\theta$ parameter.
   Its eventual observation
  would be an unambiguous  experimental  signature pointing out that: \\
  i)   a BH has been produced in the hadronic collision;\\
  ii)  there exists a minimal length inherent to short distance
  spacetime structure \cite{Garay:1994en}%,Calmet:2004mp,
  \cite{Fontanini:2005ik}.\\
  Motivated by this exciting perspective, we present
  in this paper a detailed study of the higher dimensional,
  charged, regular, mini BH. 
\section{Higher dimensional solution}
 Before we engage in a detailed calculation we would like to point out
 the basic idea of our implementation of non-commutative effects.
 It would be desirable to formulate non-commutative theories,
 including General Relativity, directly in terms of non-commuting
 coordinates. This approach would correspond to a new, deeper, level
 of ``quantization'' acting on the spacetime manifold itself rather
 than on fields, including the metric, which are structures assigned
 over the manifold itself. Unfortunately a framework of this type is not
 presently available.\\
 In a non-commutative geometry, familiar concepts loose their very meaning.
 As an example, it is worth to remark a problem
 that seems to be ignored in the current literature 
 \cite{Nasseri:2005ji}%,Nasseri:2005yr,Kar:2005qe,Chaichian:2007dr}
 while it should be especially evident when facing
 non-commutative extensions of General Relativity. Coordinate
 non-commutativity implies the existence of a \textit{finite}
 minimal length $\sqrt\theta$, below which the very concept of
 ``\textit{distance}'' becomes physically meaningless.
 Such a basic remark, immediately, raises the problem to
 define the line element, namely the \textit{infinitesimal} distance
 between two nearby points in Einstein gravity. One possible way around
 is to adopt Weyl-Wigner-Moyal $\ast$ product \cite{Chamseddine:2002fd},
 paying the cost of possible  violation of
 unitarity \cite{Gomis:2000zz}
 and Lorentz invariance in ordinary quantum field theory,
 or anisotropy of the Newton gravitational potential 
 \footnote{The alternative approach in terms of coordinate coherent
 states leads to a spherical symmetry preserving, short distance modification
 only of the Newton potential \cite{Gruppuso:2005yw}}\cite{Harikumar:2006xf}.
 Furthermore, star-product formulation of non-commutative quantum field
 theory can be handled only through perturbative
 expansion in the $\theta$ parameter.  At any finite order in $\theta$
 the approximated theory keeps no memory of its original non-locality
 and looks like an ordinary quantum field theory affected
 by higher order derivative self-interactions.  The original UV cut-off
 $1/\sqrt{\theta}$ is turned into a dimensional coupling constant leading
 to a non-normalizable UV behavior even worse than in ordinary quantum
 field theory.\\
 For all these reasons, we have developed an effective approach
 where non-commutativity is implemented only through a Gaussian de-localization
 of matter sources. In this way there is no problem in defining line
 element and Einstein's equations are kept unchanged. Our strategy can be
 summarized as follows:
 i) in non-commutative geometry there cannot be point-like object, because
    any physical distance cannot be smaller than a minimal position uncertainty
    of the order of $\sqrt\theta$;
 ii) this effect is implemented in  spacetime through matter
    de-localization, which by explicit calculations
    \cite{Smailagic:2003rp}%Smailagic:2003yb} 
    turns out to be of Gaussian form;
 iii) Spacetime geometry is determined through Einstein's equations with
    de-localized matter sources;
 iv) de-localization  of matter backfires at spacetime geometry giving
     regular, i.e. curvature singularity free, metrics. This is exactly
     what is expected from the existence of a minimal length.\\
  The presence of a universal short distance cut-off leads to the following
  effects: in quantum field theory it cures UV divergences
  \cite{Smailagic:2004yy}; in  General Relativity
  it cures curvature singularities \cite{Nicolini:2005vd,Nicolini:2005zi}.
  \\
In view of the above explanations, we are going to solve the resulting
  Einstein's equations,
 in higher dimensions, with a maximally localized source of energy and
 electric charge. The corresponding  distributions have a minimal
 spread Gaussian profile

\begin{eqnarray}
 \rho_{\mu_0}\left(\,r\,\right)&&=
\frac{\mu_0}{\left(\,4\pi\theta\,\right)^{m/2}}\,
\exp\left(-r^2/4\theta\,\right)\label{mdens}\ ,\\
\rho_q\left(\,r \,\right)&&=
\frac{q}{\left(\,4\pi\theta\,\right)^{m/2}}\,
\exp\left(-r^2/4\theta\,\right)
\label{cdens}
 \end{eqnarray}

  Matter and charge density in Eq.(\ref{mdens}), (\ref{cdens}) model  a
  physical source which is as  close as it is possible  to a
  ``point-like'' object.  $\mu_0$ is the  ``bare mass''  and $q$
  is the total electric charge. We remark that
  $\mu_0$ is only part of the total mass-energy of the system. 
  The Coulomb energy stored in the electric field is a second contribution
  to the total mass-energy sourcing the gravitational field.\\
  We are looking for static, spherically symmetric gravitational and
  electric fields solving the coupled field equations
  \begin{eqnarray}
&& R^\mu{}_\nu-\frac{1}{2}\, \delta^\mu{}_\nu\, R =
8\pi\, G\,\left(\, T^\mu_\nu\vert_{matt.} + T^\mu_\nu\vert_{el.}  \,\right)
\label{einst}\\
&& \frac{1}{\sqrt{-g}}\, \partial_\mu\,\left(\, \sqrt{-g}\, 
F^{\mu\nu}\, \right)=
J^\nu\ ,\quad  J^\mu\left(\,x\,\right)=
 4\pi\, \rho_q\left(\, r \,\right)\,\delta^\mu_0
  \label{source}   \\
&&F^{\mu\nu}=\delta^{0[\, \mu\,\vert}  \delta^{r\,\vert\, \nu \,]}\,
E\left(\, r\,\right)
\label{max}
\end{eqnarray}
where the Greek indices $\mu$, $\nu$, etc. run over $0\ ,1\ ,2\ ,\dots \ , m$;
$G=M^{1-m}_\ast$ is the  reduced fundamental  scale for $m\ge 4$
 i.e. $M_\ast\sim 1/\sqrt{\theta}\approx 1\, TeV$, while
 for $m=3$, $G=M^{-2}_{Pl.}$ i.e. $M_{Pl.}\sim
1/\sqrt{\theta}\approx 1.22\times 10^{16}\, TeV$.\\
$T^\mu_\nu\vert_{matt.}$ is the higher dimensional extension of the
energy-momentum tensor described in \cite{Nicolini:2005vd};
 $J^\mu$ is charge current which has $0$-component non-vanishing.
The Coulomb-like field satisfying the equation (\ref{max}) results to be
\begin{equation}
E\left(\, r\,\right)=\frac{2\Gamma(m/2)\,}{\pi^{(m-2)/2}}
\frac{q_\theta\left(\,r\,\right)}{r^{m-1} }\ ,\quad
q_\theta\left(\, r\,\right)=
\frac{q}{\Gamma(m/2)}
\gamma\left(\, \frac{m}{2}\ , \frac{r^2}{4\theta} \,\right)
\label{c1}
\end{equation}
where, $\gamma\left(\, a/b\ ; x\,\right) $ is the Euler lower Gamma function defined by
the following integral representation:

\begin{equation}
 \gamma\left(\, a/b\ ,x\,\right)\equiv
  \int_0^x \frac{du}{u}\, u^{a/b} \, e^{-u} \label{gamma}
\end{equation}

It turns out that the electric field is linearly vanishing
for $r\to 0$ independently of the number of space dimensions.
On the other hand, the large distance behavior is sensitive to
extra dimensions and falls-off as $1/r^{m-1}$.\\
In order to proceed, we insert solution (\ref{c1})
in the electromagnetic energy momentum tensor and
solve Einstein's equations  (\ref{einst} ). We find
Reissner-Nordstr\"om like form of the metric :

\begin{eqnarray}
&& ds^2_{(m+1)}= -h\left(\, r\,\right)\, dt^2 + h\left(\, r\,\right)^{-1}\, dr^2
+ r^2 d\Omega^2_{m-1}\label{ds}\nonumber\\
&& h\left(\, r\,\right)= 1
-\frac{4G \mu_0}{\pi^{(m-2)/2}\, r^{m-2}}
\gamma\left(\, \frac{m}{2}\ , \frac{r^2}{4\theta}\,\right)
 +\left(\,m-2\,\right)\label{rncbh}
\frac{4q^2 G}{\pi^{m-3}\, r^{2m-4}}\, F\left(\, r\,\right)\\
&& F(r)\equiv  \gamma ^2\left(\,\frac{m}{2}-1\ , \frac{r^2}{4\theta}\,\right)
 -  \frac{2^{(8-3m)/2}\,r^{m-2}}{(m-2)\theta^{(m-2)/2}}\,
  \gamma\left(\, \frac{m}{2}-1\ , \frac{r^2}{2\theta} \,\right)
\end{eqnarray}

The main feature of (\ref{rncbh}) is the absence of curvature singularities.
To prove this statement in a simple and painless way, we consider the
short-distance behavior of the metric. Inserting the small argument
 expansion of (\ref{gamma}) in (\ref{rncbh}), the geometry near the origin
 is given by

 \begin{eqnarray}
&&h\left(\, r\,\right)= 1 - \frac{4G\mu_0 }{m\, 2^{m-1}\pi^{(m-2)/2 } 
\theta^{m/2}}\, r^2
+0\left(\, r^4\,\right) \label{desit}\\
&&\Lambda_\theta \equiv \frac{4G\,\mu_0 }{m\, 2^{m-1}\pi^{(m-2)/2 } \theta^{m/2}}
\label{cc}
\end{eqnarray}

 One recognizes in  (\ref{desit}) deSitter spacetime, which is known to
 be singularity-free. Equation (\ref{cc}) shows as the bare mass $\mu_0$
 and the noncommutative parameter $\theta$ mix together to produce
 an \textit{effective cosmological constant} $\Lambda_\theta$. It is not
 surprising that different mass particles experience different cosmological
 constants. The deSitter vacuum represents the geometrical counterpart
 of the underlying noncommutative coordinate fluctuations which are taking place
 over a distance scale defined by $\sqrt{\theta}$. Different values of
 $\mu_0$ imply different Compton wavelength and ``resolution power''.
 Heavy particles can probe shorter distance better than light objects,
and  result to be more sensible to spacetime quantum fluctuations.
In our solution the  curvature singularity smears into a regular, fluctuating,
vacuum core which is  quasi-classically described by a deSitter geometry.   
Short-distance spacetime fuzziness, once it is implemented through our quasi-classical geometrical method,
cures curvature singularity as it was anticipated in the Introduction.\\
In order to proceed in our investigation, it is convenient to define the \textit{radial} mass-energy , 
$\mu\left(\, r\,\right) $
accounting for \textit{both} Coulomb and matter energy  contributions

  \begin{equation}
  \mu\left(\, r\,\right) \equiv \frac{2\mu_0 }{\pi^{ \frac{m-2}{2}}}
  \gamma\left(\, \frac{m}{2}\ , \frac{r^2}{4\theta}\,\right)
  +\frac{2^{\frac{8-3m}{2}}}{\pi^{m-3}}\frac{2q^2}{\theta^{ \frac{m}{2}-1 }}
  \gamma\left(\, \frac{m}{2}-1\ , \frac{r^2}{2\theta}\,\right)
\end{equation}

The total mass-energy, $M$,  measured by an asymptotically distant observer 
is given by

  \begin{equation}
  M = \lim_{r\to\infty}  \mu\left(\, r\,\right)=
  \frac{2\mu_0}{\pi^{\frac{m-2}{2}}}\,\Gamma\left(\, \frac{m}{2}\,\right)
  + \frac{2^{\frac{8-3m}{2}} \, 2q^2}{
  \pi^{m-3}\, \theta^{\frac{m-2}{2}}}\,
\Gamma\left(\,\frac{m}{2}-1\,\right)
  \end{equation}

Solution {\ref{rncbh}) is now expressed in terms of $M$ as

  \begin{eqnarray}
  h\left(\, r\,\right)&&= 1
-\frac{2MG}{r^{m-2} \Gamma\left(\,\frac{m}{2}\,\right)}\,
\gamma\left(\,\frac{m}{2}\ , \frac{r^2}{4\theta}\,\right)\nonumber\\
&&+ \left(\, m-2\,\right)\frac{4G\,q^2}{\pi^{m-3}\, r^{2m-4}}\,
  \left[\, F\left(\, r\,\right) + d_m\,r^{m-2}\,
  \gamma\left(\,  \frac{m}{2}\ ,  \frac{r^2}{4\theta}\, \right)\,\right]
  \label{ncrn} \\
  && d_m \equiv \frac{2^{\frac{8-3m}{2}} }{m-2}
  \, \frac{1}{\theta^{ \frac{m-2}{2} }}\,
  \frac{\Gamma\left(\, \frac{m}{2}-1\,\right)}
  {\Gamma\left(\, \frac{m}{2}\,\right)}\label{met}
  \end{eqnarray}

As one might expect, at distance $r>>\sqrt\theta$, Equation(\ref{rncbh}) gives
ordinary m-dimensional Reissner-Nordstr\"om metric

\begin{equation}
h\left(\, r\,\right)\longrightarrow
  1 -\frac{2MG}{r^{m-2}}\,
+ \left(\, m-2\,\right)\frac{4G\,q^2}{\pi^{m-3}\, r^{2m-4}}\,\Gamma^2\left(\,
\frac{m}{2}-1\,\right)
\end{equation}

At this point, it is important to address the problem of the existence of
event horizons. Their  position is determined by the roots of
the equation $h\left(\, r_+\,\right)=0$. In our case
horizon  equation cannot be solved analytically, but a plot of
$h\left(\, r\,\right)$ as a function of $r$, for fixed values of $M$, $q$, $m$
determines the positions of  horizon(s).

\begin{figure}
\begin{center}
\includegraphics[width=5cm,angle=270]{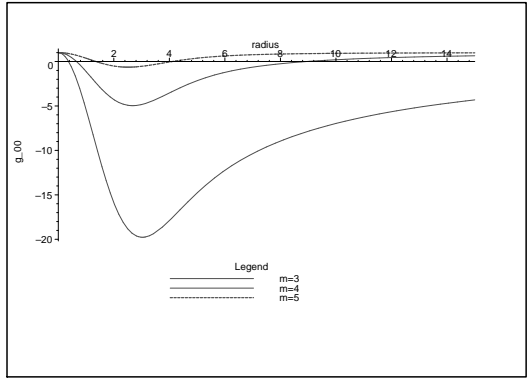}
\caption{\label{mg00} \ {\it $h\left(\, r\,\right)$ versus  $r/\sqrt{\theta}$, 
for a charge  $Q\equiv q\, M_\ast^{(m-1)/2}=\theta^{(m-2)/2} $ and
$M=40\, M_\ast $. We can observe that the inner radius decreases with 
$m$, while the outer horizon radius increases 
with $m$.}}
\end{center}
\end{figure}
%From figure (\ref{g003d}) one sees that
For any $m$ there can be two horizons 
when $M>M_0$, one degenerate horizon
for $M=M_0$ , or no horizon if $M<M_0$. $M_0$ is the mass of an 
\textit{extremal} BH and represents its final
state at the end of Hawking evaporation process. The details will be given in 
next Section.  Furthermore, it
results that increasing $m$, more and more mass is needed to create
a  black hole of a given radius. \\
The global properties of the analytically
extended solution across the horizon (coordinate singularities)
can be obtained by gluing together outer Reissner-Nordstr\"om and
inner deSitter patches.
On general ground horizons are solutions of the equation
 
\begin{eqnarray}
    &&M=U\left(\, r_H\ ; q\,\right)\\
    && U\left(\, r_H\ ; q\,\right)=
    2^{\frac{8m-3}{2}}\,\Gamma\left(\,\frac{m-2}{2}\,\right)
    \frac{2q^2}{ \pi^{m-3}}\frac{1}{\theta^{\frac{m-2}{2}}}\nonumber\\
  &&+\frac{ \Gamma\left(\,\frac{m}{2}\,\right)}
  {2G\gamma\left(\, m/2\ ,r^2_\pm /4\theta\,\right)}\left[\, r_H^{m-2}
  +\frac{(m-2)}{\pi^{m-3}} \frac{4G\,q^2}{ r_H^{m-2} }\,
  F\left(\,r_H \,\right)\,
  \right]\label{plot}
 \end{eqnarray}

\begin{figure}[ht!]
\begin{center}
\includegraphics[width=5cm,angle=270]{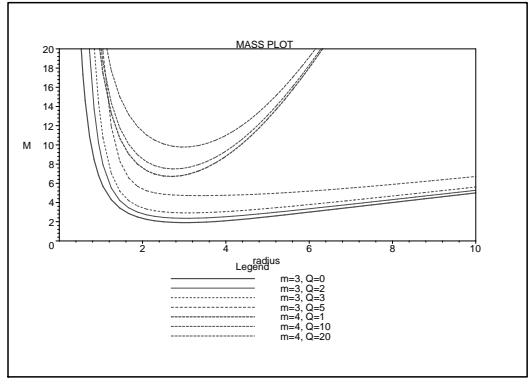}
\caption{\label{massa0}
\textit{$U\left(\, r_H\ ;q\,\right)$  for different $q$ and $m$.}}
\end{center}
\end{figure}
Eq.(\ref{plot}) offers an alternative way of studying existence of
horizons. \\
 The use of the equation relating
  the total mass energy of the system to the radius of the event
  horizon follows the approach  proposed in
  \cite{Aurilia:1984cm}%,Aurilia:1987cp,Blau:1986cw,Farhi:1989yr}
   with the advantage of allowing
   an in-depth investigation of geometry and dynamics of the system.
  We see that the sole effect of extra-dimensions
   is to lift upward the $3+1$ dimensional curve thus increasing
   the value of the mass for a given radius of the horizon, including
   the degenerate case.  Thus, we conclude that the evolution of the
   BH towards its extremal configuration is qualitatively the same as
   described in \cite{Nicolini:2005vd}. The increase of the mass $M_0$
   and its eventual experimental verification could indicate
   the number of extra-dimensions.
   \section{Black hole thermodynamics}
In this section we are going to investigate some thermodynamic
properties of the regular BH described by the line element
(\ref{rncbh}). The starting point is
the explicit form of the  Hawking temperature $T_H$:

\begin{eqnarray}
&& 4\pi\,T_H = \frac{1}{r_+}\,\left[\, m - 2 - r_+
 \frac{\gamma^\prime\left(\, m/2\ ,r_ +^2 /4\theta\,\right)}
 {\gamma\left(\, m/2\ , r_ +  ^2 /4\theta \,\right)}\,\right]+\nonumber\\
 &&-\frac{16\,G\,q^2 }
{ \pi ^{m - 3} r_ + ^{2m - 3}} \left[\, \gamma ^2\left(\,
m/2\ , r_ +  ^2 /4\theta \,\right)
 +\left(\,m-2\,\right)\frac{r_+}{4}\, F\left(\, r_+ \,\right)\,
 \frac{\gamma^\prime\left(\, m/2\ ,r_ + ^2 /4\theta\,
 \right)}{\gamma \left(\, m/2\ , r_ +  ^2 /4\theta\, \right)}\right]\nonumber\\
&&\label{temp}
\end{eqnarray}

where, we replaced $M$ with $r_+$ by using the horizon equation (\ref{plot}).
\begin{figure}[ht]
\begin{center}
\includegraphics[width=5cm,angle=270]{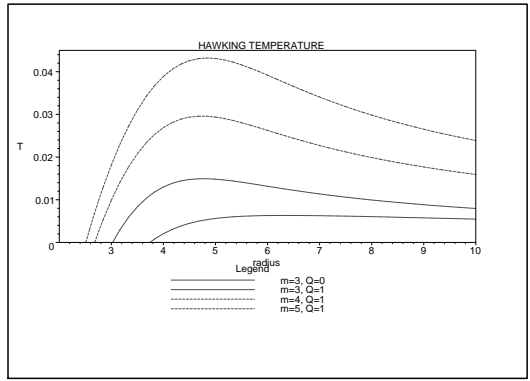}
\caption{ \label{bhtemp}
\textit{Hawking temperature  $T_H \sqrt\theta$  as a function of
 $r_+/\sqrt\theta$, for different values of $Q/{\sqrt\theta}^{(m-2)}$ and $m$.
 $T_H$ reaches a maximum whose value increases with $m$ before dropping
 to zero.}}
\end{center}
\end{figure}
From the plot of the temperature one sees that in any dimension there is
an initial ``~\textit{Schwarzschild-phase}~'' for $r_+ \geq 7\,\sqrt\theta$.
Approaching maximum temperature, a departure from the Schwarzschild behavior
shows up.  The maximum temperature occurs in the range $ 4.5\,\sqrt\theta
\le r_+ \leq 7\,\sqrt\theta $. Immediately after the temperature drops down
and the BH enters a \textit{nearly-extremal} phase, asymptotically approaching
a zero-temperature, degenerate configuration.
This is the same behavior already encountered in the neutral
case \cite{Nicolini:2005vd}.
The effect of extra-dimensions is to shift upwards
the temperature while shrinking the radius of the extremal BH.
The effect of charge is just to lower the  temperature as can be
seen from the two curves with   $m=3$ and $Q=0\ , 1$.
Therefore, upper bounds for the maximal temperature can be obtained from
the plot of $T_H$ in the neutral case. \\
One can see from  (\ref{temp}) that the peak temperature increases with $m$, 
but even in the case  $m=10$ it is
about $98 \, GeV\,\left(\simeq 10^{15}\,K\right)$ which is much lower than 
$M_\ast$.  
For $m=3$ i.e. no extra-dimensions, the fundamental
scale $M_\ast=M_{Pl.}$. Notwithstanding, back-reaction is still negligible as 
the maximal temperature is $18\times
10^{16}\, GeV$ much lower than $T_{Pl.}$.
\begin{center}
  \begin{table}[htb]
 \caption{ BH remnant (minimal) masses and  (minimal)
 radii for different values of $m$}
 {\begin{tabular}{@{}ccccccccccccccccc@{}} \hline\hline 
 m
& & 3 & &  5 & & 6 & &  9 & & 10 \\
 \hline 
 $M_0$ (TeV)
& & 2.3 $\times 10^{16} $ & &  $24$ & & $94$ & & 
 $7.3\times 10^3$ & & $3.4\times 10^4$ \\
\hline 
$r_0$ ($10^{-4}$ fm)
& & $4.88 \times 10^{-16} $  & &  $4.95$ & & $4.75$ & & 
 $4.46$ & & $4.40$ \\
\hline\hline
\end{tabular} \label{ta2}}
\end{table}
\end{center}
The energy of the electrostatic field increases the total mass-energy. Thus, 
the minimal value for $M_0$ as a
function of $m$ can be obtained studying the neutral case $Q=0$.  This is given 
in Table (\ref{ta2}), where we
also listed the corresponding radius of the event horizon. The BH production 
cross section is simply the area of
the event horizon, which leads to an estimate of the order of $10\, nb$, for 
every $m$. This is ``good news''. On
the other hand, the ``bad news'' is that for $m\ge 6$ the remnant is too heavy 
to be produced at LHC
\cite{Rizzo:2006zb}. Maybe, these kind of objects could be detected in 
Ultra-High-Energy cosmic rays
\cite{Nagano:2000ve,Cavaglia:2002si}. \\

In this section we are also going to investigate
 eventual effect of the minimal length on the expression of the BH entropy
 \cite{Myung:2006mz}%,Myung:2007qt,Myung:2007xd,Myung:2007av},
 in terms of the area of the event horizon.\\
  It can be useful to recall
 that the celebrated area/entropy law was originally derived by Bekenstein
 in the framework of information theory. The correct proportionality constant
 between entropy and  area of the event horizon was later obtained by 
 Hawking through a thermodynamic approach. It can be worth to remark that
 this relation holds for classical \footnote{Quantum effects introduce
 logarithmic corrections.}, black, point-like, singular, objects, and there is 
 no \textit{a priori} evidence that it would keep the same form in our case. We
 must derive it ``from scratch'', so to say. Under this respect,
 the thermodynamic approach is easier to carry out.  We start from
 the fundamental law of black hole thermodynamics 
 $dM = T_H\, dS_H + \phi_H\, dq$
where, $\phi_H\equiv\phi\left(\, r_+\,\right)$ is the electrostatic potential
on the event horizon.
As physical quantities  are evaluated on the event horizon,
$M$ can be expressed in terms of $U\left(\, r_+\ ; q\,\right)$ and $dM$
can be written as
\begin{equation}
dM= \frac{\partial U}{\partial r_+ }\, d r_+
+ \frac{\partial U}{\partial q}\, dq
\label{M2}
\end{equation}
From the two different forms of $dM$ one finds 
the following expression for the BH entropy

\begin{equation}
dS_H= \frac{1}{T_H}\, \frac{\partial U}{\partial r_+}\,d r_+
\label{dsh}
\end{equation}

We start from (\ref{plot}) and calculate $\partial U/\partial r_+$

\begin{eqnarray}
 &&\frac{\partial U}{\partial r_+} 
 = \frac{\Gamma\left(\,m/2\,\right)\,r^{m - 3}_+}{2G\gamma\left(\, m/2\ ,
 r^2_+/4\theta\,\right)}
 \left\{\, m - 2 - r_+\,\frac{\gamma^\prime\left(\, m/2\ ,r^2_+ /4\theta\,\right)}
 {\gamma\left(\, m/2\ ,r^2_+ /4\theta \,\right)}+ \right.\nonumber\\
 && -\frac{16G\,q^2 }{\pi^{m - 3}\, r^{2m - 4}_+}\left.\left[\,
 \gamma ^2 \left(\,m/2\ ,r^2_+ /4\theta\,\right)+\left(\,m - 2\,\right)
  \frac{r_+}{4}\,
 F\left(\, r_+\,\right)\,\frac{\gamma^\prime\left(\, m/2\ ,r^2_+ /4\theta 
 \,\right)}
 {\gamma\left(\, m/2\ ,r^2_+ /4\theta\,\right)}\,\right]\,\right\}
 \nonumber
% &&\label{mss}
\end{eqnarray}
$\partial U/ \partial r_+ $ is confronted with the expression 
(\ref{temp}) for the BH temperature  leading  to the relation

  \begin{equation}
\frac{\partial U}{\partial r_+}= 2\pi\, r_+^{m-2}\,
\frac{\Gamma\left(\,m/2\,\right)}
{\gamma\left(\, m/2\ , r_+^2/4\theta\,\right) }\, T_H
\end{equation}

The final expression for  the BH entropy variation is

\begin{equation}
dS_H=2\pi\, r_+^{m-2}\,
\frac{\Gamma\left(\,m/2\,\right)}
{\gamma\left(\, m/2\ , r_+^2/4\theta\,\right) }\, dr_+
\label{dsh2}
\end{equation}

In order to obtain the BH entropy we integrate (\ref{dsh2}) from the extremal
horizon $r_e$ up to a generic external horizon $r_+$. The result is

\begin{eqnarray}
\Delta S_H=&& \frac{2\pi}{m-1} \Gamma\left(\, m/2\,\right)\,\left[\,
\frac{r_+^{m-1}}{\gamma\left(\, m/2\ , r_+^2/4\theta\,\right)}
-\frac{r_e^{m-1}}{\gamma\left(\, m/2\ , r_e^2/4\theta\,\right)}\,\right]
+\nonumber\\
+&& \frac{2\pi}{m-1} \Gamma\left(\, m/2\,\right)\,
\int_{r_e}^{r_+} dx\,x^{m-1} \frac{\gamma^\prime\left(\, m/2\ ,x^2/4\theta\,
\right)}{\gamma\left(\, m/2\ ,x^2/4\theta\,\right)^2}
\label{sbh}
\end{eqnarray}

Taking into account that the area of the event horizon is
$ A_H= \frac{2\, \pi^{m/2}}{\Gamma\left(\, m/2\,\right)}\, r_+^{m-1} $
we rewrite (\ref{sbh}) as

\begin{eqnarray}
&&\Delta S_H=\frac{\Gamma\left(\, m/2\,\right) }{\pi^{\frac{m}{2}-1}
  \left(\,m-1\,\right)}\,\frac{1}{G_\theta\left(\, r\,\right)} \,\left(\,
A_+  -A_e\,\right)+\delta S_H\label{unquarto2}\\
&& G_\theta\left(\, r_+\,\right)=
\frac{G}{\Gamma\left(\,\frac{m}{2}\,\right)}
\, \gamma\left(\, \frac{m}{2}\ , \frac{r^2_+}{4\theta}\,\right)
\end{eqnarray}

The first term in equation (\ref{unquarto2})  generalizes the celebrated four
dimensional relation $S_H= A_H/4G_N$ to higher dimensions. It is worth noticing
that the Newton constant is replaced by the effective gravitational coupling.
A similar conclusion has been recently attained, in a different framework, in
\cite{Brustein:2007jj}. \\
Finally, let us analyze in more details correction term $\delta S_H$
in (\ref{unquarto2}).
It can be seen that $r_e> \sqrt\theta$ always,
so we can approximate $\gamma\left(\, m/2\ , r_e^2/4\theta\,\right)\approx
\Gamma\left(\, m/2\,\right)$ and write ( in Planck units $G=1$)

\begin{equation}
\delta S_H
\approx \frac{1}{2^{m-2}}\frac{1}{\theta^{\frac{m}{2}-1}}\left(\,
r_e^{2m-3} e^{-r_e^2/4\theta} -r_+^{2m-3} e^{-r_+^2/4\theta}\,\right)
\end{equation}
We conclude that the area law is maintained up to exponentially small
corrections.\\
We have shown  that the  temperature of a 
Reissner-Nordstr\"{o}m type BH never diverges.
Thus, the end-point of the Hawking evaporation is a nearly extremal BH in 
thermal equilibrium with the
environment. However, in the four dimensional picture
 this is not a satisfactory candidate to the role of remnant because
charged BHs results to be quantum mechanically
unstable under pair production.  A mini BH decays into a neutral
Schwarzschild type object much before reaching a significant temperature.
Once in the Schwarzschild phase the temperature will increase without
limit leading to an unpredictable final stage.  This is not a problem
in our case thanks to the presence of the $\theta$ parameter even in
neutral phase. 
\section{Pair creation}
In this section we are going to study the discharge process with
a special attention to the way the presence of extra dimension
affects the mean life of the charged object.\\
The creation of $e^\pm$ pairs near the event horizon is
  described by the Schwinger formula \cite{Dunne:2004nc}.
  This formula implies
  that in order for creation process to take place, the electric field
  has to exceed the \textit{critical intensity} $E_c\equiv m^2_e/e$.
   Over-criticality condition leads to

  \begin{equation}
  \frac{2q}{\pi^{\frac{m-2}{2}}\, r^{m-1}}\,
\gamma\left(\, \frac{m}{2}\ , \frac{r^2}{4\theta}\,\right)
\ge  \frac{m^2_e}{e} \label{schw}
\end{equation}

The concept of  Dyadosphere 
has been introduced in \cite{Preparata:1998rz,Ruffini:1998df} as  
the spherical region, of radius $r_{ds}$, where (\ref{schw}) is valid.
Recently, there has a criticism about the possibility
of Dyadosphere development for astrophysical objects
\cite{Page:2006cm}. %,Page:2006zk,Page:2006zi}
Without going into this debate, we would like to notice
that in the case of micro BH the condition for the
existence of the  Dyadosphere is always met in any dimension.
In our case, the radius $r_{ds}$ is determined by

\begin{equation}
r_{ds}^{m-1}= \frac{2e q}{\pi^{(m-2)/2}\, m^2_e}\,
\gamma\left(\,\frac{m}{2}\ ,\frac{r_{ds}^2}{4 \theta }\,\right)
\label{rdyado}
\end{equation}
Equation  (\ref{rdyado}) can be solved numerically. However,
in what follows we do not need the explicit value of $r_{ds} $.
%\begin{equation}
% r_{ds}^{m-1}\simeq \frac{2 eq}{\pi^{(m-2)/2}\,  m^2_e}\,
%\gamma\left(\,\frac{m}{2}\ ,\frac{r_0^2 }{4\theta}\,\right)\ ,\quad
% r_0^{m-1}\equiv \frac{2 eq}{\pi^{(m-2)/2}\,  m^2_e}\label{dyado}
%\end{equation}
In order to proceed, we introduce a surface charge density as:

\begin{equation}
\sigma\left(\,r\,\right)= \frac{q_\theta\left(\,r\,\right)}{A}=
q_\theta\left(\,r\,\right)
\frac{\,\Gamma\left(\,\frac{m}{2}\,\right)}{ 2\pi^{m/2}r^{m-1} }
\end{equation}

The idea is to divide the Dyadosphere into ``thin'' spherical
shells of thickness equal to the electron Compton wavelength
$\lambda_e$. Within each shell the electric field can be considered
constant and described by
$E\left(\,r\,\right)= 4\pi\, \sigma\left(\,r\,\right)$.
Such a description in terms of constant field is necessary in order to
apply the Schwinger formula.\\
The \textit{critical surface density} is obtained when $E=E_c$:
It can be inferred from  Fig.(\ref{mg00}) and the definition
(\ref{rdyado}) that $r_{ds}> r_+\ge r_0> \sqrt\theta$. Thus,
we see that the dependence of the  critical density from $m$ is
confined only within the electric charge.
For $\sigma\ge\sigma_c $ $e_\pm$-pairs are created and
the rate of their production is given by

\begin{equation}
W=\frac{1}{2^{m+1}\pi^m}\left[\, 4\pi\, e\, \sigma\right]^{(m+1)/2}
\exp\left(-\frac{\, m^2_e}{4e \,\sigma}\right)
\end{equation}

The total number of pairs produced, in one second, inside a spherical
shell of thickness $\lambda_e= 1/2m_e$  is
\begin{equation}
\frac{\Delta N}{\Delta \tau}\equiv \lambda_e\, A\left(\,r\,\right)\,
  W=\frac{  \lambda_e \, r^{m-1}  m_e^{m+1} }{2^m \, \pi^{m/2}\,
\Gamma\left(\,\frac{m}{2}\,\right) }\left(\,\frac{\sigma}
 {\sigma_c}\, \right)^{(m+1)/2}
\exp\left(-\pi\,\frac{\sigma_c}{\sigma}\right)
\end{equation}
 The discharging process is taking place until
the critical density is reached. Then, it becomes
exponentially suppressed. The decay time can be obtained by noticing that

\begin{equation}
\sigma -\sigma_c = \frac{e}{A\left(\,r\,\right) }\frac{\Delta N}{\Delta \tau}
\Delta \tau=e\,\lambda_e\,W\,\Delta\tau
\end{equation}

Thus, we  estimate the discharge time to be

\begin{equation}
 \Delta \tau = \left(\frac{2\pi}{m_e}\right)^{m-1}\frac{1}{e^2\,\lambda_e}\,
\frac{x -1}{x^{(m+1)/2}}
\exp\left(\,\frac{\pi}{x}\,\right)
\label{meanlife2}
\end{equation}
 where, we introduced the variable $x=\sigma/\sigma_c$. 
 In order to restore the dependence from the minimal length let us recall
 that the electric charge is a dimensional quantity. Thus, we write
$e^2=4\pi\alpha_{em}\,L^{m-3}$
where, $\alpha_{em}=1/137 $ is the fine structure constant in four dimensions,
and $L$ is a characteristic length scale of the higher dimensional theory.
In our approach there is only one length scale and it is natural to identify
$L=\sqrt{\theta}$. Then, the discharge mean time turns out to be
(in c.g.s. units )

\begin{equation}
 \Delta \tau = \frac{\theta\, m_e}{\alpha_{em}}\,
\left(\frac{2\pi}{m_e c\sqrt{\theta}}\right)^{m-1}
\frac{x -1}{x^{(m+1)/2}}
\exp\left(\,\frac{\pi}{x}\,\right)
\label{meanlifem}
\end{equation}

 where we estimate $m_e\, c\, \sqrt\theta \approx 0.5\times 10^{-6}$, which is
 compatible with expected length scale in $TeV$ quantum gravity.
 Eq.(\ref{meanlifem}) gives the discharge time assuming that the process
 takes place in the $m+1$ dimensional bulk spacetime. It is interesting
 to compare it with corresponding expression for $m=3$ which describes
 pair creation on the ``three-brane'' world. Thus we find

 \begin{equation}
\frac{\Delta \tau }{\Delta \tau_{m=3} }=
\left(\,\frac{2\pi}{m_e c\sqrt\theta}\,\right)^{m-3}
\frac{1}{x^{(m-3)/2}}
\end{equation}

An upper bound on $\Delta \tau $ can be obtained for $x\approx 1$, i.e.
 $\sigma \approx \sigma_c$, and $\Delta \tau_{m=3}\le 1.76\times 10^{-19}sec$
 \cite{Ruffini:1998df}:

 \begin{equation}
 \Delta \tau\le \left(\,\frac{2\pi}{m_e c\sqrt\theta}\,\right)^{m-3}
 1.76\times 10^{-19}sec \label{tau}
 \end{equation}

 As there are no evidence of  the presence of a minimal length at the
 atomic length scale,  $m_e c\sqrt\theta<<1$.  Eq.(\ref{tau}) shows that
 both $\theta$ and
 $m$ contribute to the increase of the mean life by several orders of magnitude,
  already for  $m$ slightly greater than three. \\
  We conclude that: if the decay takes place in the bulk the discharge
  time increases significantly as a function of $m$. If, on the contrary,
  brane universe scenario, where standard model elementary particles are
  confined on the brane, is realized in nature $\Delta \tau$ should be
  of the order of $10^{-19}\, sec$.\\
  Two comments are in order. From  (\ref{tau}) we see that
  Schwinger mechanism is relevant only in the case $m\le 6$. For
  higher dimension charge neutralization will take place through
  Hawking emission, as soon as the BH temperature is higher than
  $1.22\, MeV$, and
  capture of surrounding charged particles produced in the
  hadronic collision. Secondly, our considerations assume electrons
  to be free to move in the whole higher dimensional space.
  Brane models constrain ordinary matter particle to be glued
  to the brane itself, while only gravity can probe transverse
  higher dimensions. Even if a D-brane is not explicitly present in our
  geometry, line element (\ref{ds})  can approximate a D-brane geometry
  provided the radius of the event horizon is small with respect to
  the size of the extra dimensions. To take into account models where
  gauge fields are confined to the brane, Schwinger mechanism must
  be restricted to take place only inside the brane and the relevant 
  discharge time is given by $10^{-19}\,s$.


\begin{thebibliography}{99}
\bibitem{Witten:1995im}
  E.~Witten,
  %``Bound states of strings and p-branes,''
  Nucl.\ Phys.\  B {\bf 460}, 335 (1996)
  [arXiv:hep-th/9510135]\\
%\bibitem{Seiberg:1999vs}
  N.~Seiberg and E.~Witten,
  %``String theory and noncommutative geometry,''
  JHEP {\bf 9909}, 032 (1999)
  [arXiv:hep-th/9908142]
�������%\cite{Snyder:1946qz}
\bibitem{Snyder:1946qz}
  H.~S.~Snyder,
  %``Quantized space-time,''
  Phys.\ Rev.\  {\bf 71}, 38 (1947)
\bibitem{ArkaniHamed:1998rs}
  N.~Arkani-Hamed, S.~Dimopoulos and G.~R.~Dvali,
  %``The hierarchy problem and new dimensions at a millimeter,''
  Phys.\ Lett.\  B {\bf 429}, 263 (1998)
  [arXiv:hep-ph/9803315]\\
%\bibitem{Antoniadis:1998ig}
  I.~Antoniadis, N.~Arkani-Hamed, S.~Dimopoulos and G.~R.~Dvali,
  %``New dimensions at a millimeter to a Fermi and superstrings at a TeV,''
  Phys.\ Lett.\  B {\bf 436}, 257 (1998)
  [arXiv:hep-ph/9804398]
\bibitem{Giddings:2001ih}
  S.~B.~Giddings,
  %``Black hole production in TeV-scale gravity, and the future of high  energy
  %physics,''
in {\it Proc. of the APS/DPF/DPB Summer Study on the Future of Particle Physics
 (Snowmass 2001) } ed. N.~Graf,
{\it In the Proceedings of APS / DPF / DPB Summer Study on the Future of 
Particle Physics (Snowmass 2001), Snowmass, Colorado, 30 Jun - 21 Jul
2001, pp P328}
  [arXiv:hep-ph/0110127]\\
%\bibitem{Bleicher:2001kh}
  M.~Bleicher, S.~Hofmann, S.~Hossenfelder and H.~Stoecker,
  %``Black hole production in large extra dimensions at the Tevatron: A  chance
  %to observe a first glimpse of TeV scale gravity,''
  Phys.\ Lett.\  B {\bf 548}, 73 (2002)
  [arXiv:hep-ph/0112186]\\
%\bibitem{Stenmark:2002yb}
  M.~Stenmark,
  %``Decay Of Tev Scale Black Holes And Massive Continuum Kaluza-Klein Towers''
  Chin.\ J.\ Phys.\  {\bf 40} (2002) 512\\
%\cite{Hsu:2002bd}
%\bibitem{Hsu:2002bd}
  S.~D.~H.~Hsu,
  %``Quantum production of black holes,''
  Phys.\ Lett.\  B {\bf 555}, 92 (2003)
  [arXiv:hep-ph/0203154]\\
  %%CITATION = PHLTA,B555,92;%%
%\bibitem{Cavaglia:2003qk}
  M.~Cavaglia, S.~Das and R.~Maartens,
  %``Will we observe black holes at LHC?,''
  Class.\ Quant.\ Grav.\  {\bf 20}, L205 (2003)
  [arXiv:hep-ph/0305223]\\
%\cite{Chamblin:2004zg}
%\bibitem{Chamblin:2004zg}
  A.~Chamblin, F.~Cooper and G.~C.~Nayak,
  %``SUSY production from TeV scale blackhole at LHC,''
  Phys.\ Rev.\  D {\bf 70}, 075018 (2004)
  [arXiv:hep-ph/0405054]\\
%\bibitem{Cavaglia:2004jw}
  M.~Cavaglia and S.~Das,
  %``How classical are TeV-scale black holes?,''
  Class.\ Quant.\ Grav.\  {\bf 21}, 4511 (2004)
  [arXiv:hep-th/0404050]\\
%\bibitem{Shankaranarayanan:2003qm}
  S.~Shankaranarayanan and N.~Dadhich,
  %``Non-singular black-holes on the brane,''
  Int.\ J.\ Mod.\ Phys.\  D {\bf 13}, 1095 (2004)\\
 %\bibitem{Rizzo:2006uz}
  T.~G.~Rizzo,
  %``TeV-scale black hole lifetimes in extra-dimensional Lovelock gravity,''
  Class.\ Quant.\ Grav.\  {\bf 23}, 4263 (2006)
  [arXiv:hep-ph/0601029].\\
%\bibitem{Rizzo:2005jz}
  T.~G.~Rizzo,
  ``TeV-scale black holes in warped higher-curvature gravity''
  arXiv:hep-ph/0510420.
\bibitem{Casanova:2005id}
  A.~Casanova and E.~Spallucci,
  %``TeV mini black hole decay at future colliders,''
  Class.\ Quant.\ Grav.\  {\bf 23}, R45 (2006)
  [arXiv:hep-ph/0512063].
\bibitem{Alberghi:2006qr}
  G.~L.~Alberghi, R.~Casadio, D.~Galli, D.~Gregori, A.~Tronconi and V.~Vagnoni,
  %``Probing quantum gravity effects in black holes at LHC,''
  arXiv:hep-ph/0601243\\
%\bibitem{Casadio:2001wh}
  R.~Casadio and B.~Harms,
  %``Can black holes and naked singularities be detected in accelerators?,''
  Int.\ J.\ Mod.\ Phys.\  A {\bf 17}, 4635 (2002)
  [arXiv:hep-th/0110255].
 \bibitem{Stoecker:2006yz}
  H.~Stoecker,
  %``Mini black holes in the first year of the LHC: Discovery through di-jet
  %suppression, multiple mono-jet emission and ionizing tracks in ALICE,''
  J.\ Phys.\ G {\bf 32}, S429 (2006).
\bibitem{Nasseri:2005ji}
  F.~Nasseri,
  %``Schwarzschild black hole in noncommutative spaces,''
  Gen.\ Rel.\ Grav.\  {\bf 37}, 2223 (2005)
  [arXiv:hep-th/0508051]\\
%\bibitem{Nasseri:2005yr}
  F.~Nasseri,
  %``Event horizon of Schwarzschild black hole in noncommutative spaces,''
  Int.\ J.\ Mod.\ Phys.\  D {\bf 15}, 1113 (2006)
  [arXiv:hep-th/0508122]\\
%\bibitem{Kar:2005qe}
  S.~Kar and S.~Majumdar,
  %``Black hole geometries in noncommutative string theory,''
  Int.\ J.\ Mod.\ Phys.\  A {\bf 21}, 6087 (2006)
  [arXiv:hep-th/0510043]\\
%\bibitem{Chaichian:2007dr}
  M.~Chaichian, M.~R.~Setare, A.~Tureanu and G.~Zet,
  ``On Black Holes and Cosmological Constant in Noncommutative Gauge Theory of
  Gravity,''
  arXiv:0711.4546 [hep-th]\\
%\bibitem{Chaichian:2007we}
  M.~Chaichian, A.~Tureanu and G.~Zet,
  ``Corrections to Schwarzschild Solution in Noncommutative Gauge Theory of
  Gravity,''
  arXiv:0710.2075 [hep-th]
%\cite{Kobakhidze:2007jn}
\bibitem{Kobakhidze:2007jn}
  A.~Kobakhidze,
  ``Noncommutative corrections to classical black holes,''
  arXiv:0712.0642 [gr-qc]\\
  %%CITATION = ARXIV:0712.0642;%%
%\bibitem{Mukherjee:2007fa}
  P.~Mukherjee and A.~Saha,
  %``Reissner--Nordstrom solutions in noncommutative gravity,''
  arXiv:0710.5847 [hep-th]
%\cite{Nicolini:2005vd}
\bibitem{Nicolini:2005vd}
  P.~Nicolini, A.~Smailagic and E.~Spallucci,
  %``Noncommutative geometry inspired Schwarzschild black hole,''
  Phys.\ Lett.\  B {\bf 632}, 547 (2006)
  [arXiv:gr-qc/0510112]\\
  %%CITATION = PHLTA,B632,547;%%
 %\cite{Ansoldi:2006vg}
%\bibitem{Ansoldi:2006vg}
  S.~Ansoldi, P.~Nicolini, A.~Smailagic and E.~Spallucci,
  %``Noncommutative geometry inspired charged black holes,''
  Phys.\ Lett.\  B {\bf 645}, 261 (2007)
  [arXiv:gr-qc/0612035]
  %%CITATION = PHLTA,B645,261;%%
\bibitem{Rizzo:2006zb}
  T.~G.~Rizzo,
  %``Noncommutative inspired black holes in extra dimensions,''
  JHEP {\bf 0609}, 021 (2006)
  [arXiv:hep-ph/0606051]
\bibitem{Nagano:2000ve}
  M.~Nagano and A.~A.~Watson,
  %``Observations And Implications Of The Ultrahigh-Energy Cosmic Rays,''
  Rev.\ Mod.\ Phys.\  {\bf 72}, 689 (2000).
\bibitem{Cavaglia:2002si}
  M.~Cavaglia,
  %``Black hole and brane production in TeV gravity: A review,''
  Int.\ J.\ Mod.\ Phys.\  A {\bf 18}, 1843 (2003)
  [arXiv:hep-ph/0210296]
\bibitem{Garay:1994en}
  L.~J.~Garay,
  %``Quantum gravity and minimum length,''
  Int.\ J.\ Mod.\ Phys.\  A {\bf 10}, 145 (1995)
  [arXiv:gr-qc/9403008]\\
%\cite{Calmet:2004mp}
%\bibitem{Calmet:2004mp}
  X.~Calmet, M.~Graesser and S.~D.~H.~Hsu,
  %``Minimum length from quantum mechanics and general relativity,''
  Phys.\ Rev.\ Lett.\  {\bf 93}, 211101 (2004)
  [arXiv:hep-th/0405033]
\bibitem{Fontanini:2005ik}
  M.~Fontanini, E.~Spallucci and T.~Padmanabhan,
  %``Zero-point length from string fluctuations,''
  Phys.\ Lett.\  B {\bf 633}, 627 (2006)
  [arXiv:hep-th/0509090]
\bibitem{Chamseddine:2002fd}
  A.~H.~Chamseddine,
  %``Invariant Actions For Noncommutative Gravity,''
  J.\ Math.\ Phys.\  {\bf 44}, 2534 (2003)
  [arXiv:hep-th/0202137]
\bibitem{Gomis:2000zz}
J.~Gomis and T.~Mehen,
  %``Space-time noncommutative field theories and unitarity,''
  Nucl.\ Phys.\  B {\bf 591}, 265 (2000)
  [arXiv:hep-th/0005129].
\bibitem{Harikumar:2006xf}
  E.~Harikumar and V.~O.~Rivelles,
  %``Noncommutative gravity,''
  Class.\ Quant.\ Grav.\  {\bf 23}, 7551 (2006)
  [arXiv:hep-th/0607115]
\bibitem{Gruppuso:2005yw}
  A.~Gruppuso,
  %``Newton's law in an effective non commutative space-time,''
  J.\ Phys.\ A  {\bf 38}, 2039 (2005)
  [arXiv:hep-th/0502144]
%\cite{Smailagic:2003rp}
\bibitem{Smailagic:2003rp}
  A.~Smailagic and E.~Spallucci,
  %``UV divergence-free QFT on noncommutative plane,''
  J.\ Phys.\ A  {\bf 36}, L517 (2003)
  [arXiv:hep-th/0308193]
  %%CITATION = JPAGB,A36,L517;%%
 %\cite{Smailagic:2003yb}
%\bibitem{Smailagic:2003yb}
  A.~Smailagic and E.~Spallucci,
  %``Feynman path integral on the noncommutative plane,''
  J.\ Phys.\ A  {\bf 36}, L467 (2003)
  [arXiv:hep-th/0307217]
  %%CITATION = JPAGB,A36,L467;%%
\bibitem{Smailagic:2004yy}
  A.~Smailagic and E.~Spallucci,
  %``Lorentz invariance and unitarity in UV-finiteness of QFT 
  %on  noncommutative
  %spacetime,''
  J.\ Phys.\ A  {\bf 37}, 1 (2004)
  [Erratum-ibid.\  A {\bf 37}, 7169 (2004)]
  [arXiv:hep-th/0406174]
%\cite{Nicolini:2005zi}
\bibitem{Nicolini:2005zi}
  P.~Nicolini,
  %``A model of radiating black hole in noncommutative geometry,''
  J.\ Phys.\ A  {\bf 38}, L631 (2005)
  [arXiv:hep-th/0507266]
  \bibitem{Aurilia:1984cm}
  A.~Aurilia, G.~Denardo, F.~Legovini and E.~Spallucci,
  %``Vacuum Tension Effects On The Evolution Of Domain Walls In The Early
  %Universe,''
  Nucl.\ Phys.\  B {\bf 252}, 523 (1985)\\
  %\bibitem{Aurilia:1987cp}
  A.~Aurilia, R.~S.~Kissack, R.~Mann and E.~Spallucci,
  %``RELATIVISTIC BUBBLE DYNAMICS: FROM COSMIC INFLATION TO HADRONIC BAGS,''
  Phys.\ Rev.\  D {\bf 35}, 2961 (1987)\\
  %\bibitem{Blau:1986cw}
  S.~K.~Blau, E.~I.~Guendelman and A.~H.~Guth,
  %``The Dynamics of False Vacuum Bubbles,''
  Phys.\ Rev.\  D {\bf 35}, 1747 (1987)\\
 % \bibitem{Farhi:1989yr}
  E.~Farhi, A.~H.~Guth and J.~Guven,
  %``IS IT POSSIBLE TO CREATE A UNIVERSE IN THE LABORATORY BY QUANTUM
  %TUNNELING?,''
  Nucl.\ Phys.\  B {\bf 339}, 417 (1990)
\bibitem{Myung:2006mz}
  Y.~S.~Myung, Y.~W.~Kim and Y.~J.~Park,
  %``Thermodynamics and evaporation of the noncommutative black hole,''
  JHEP {\bf 0702}, 012 (2007)
  [arXiv:gr-qc/0611130]\\
  %%CITATION = JHEPA,0702,012;%%
 %\cite{Myung:2007qt}
%\bibitem{Myung:2007qt}
  Y.~S.~Myung, Y.~W.~Kim and Y.~J.~Park,
  %``Quantum Cooling Evaporation Process in Regular Black Holes,''
  Phys.\ Lett.\  B {\bf 656}, 221 (2007)
  [arXiv:gr-qc/0702145]\\
  %%CITATION = PHLTA,B656,221;%%
  %\cite{Myung:2007xd}
%\bibitem{Myung:2007xd}
  Y.~S.~Myung, Y.~W.~Kim and Y.~J.~Park,
  ``Entropy of an extremal regular black hole''
  arXiv:0705.2478 [gr-qc]\\
%\cite{Myung:2007av}
%\bibitem{Myung:2007av}
  Y.~S.~Myung, Y.~W.~Kim and Y.~J.~Park,
  ``Thermodynamics of regular black hole''
  arXiv:0708.3145 [gr-qc]
  \bibitem{Brustein:2007jj}
  R.~Brustein, D.~Gorbonos and M.~Hadad,
  ``Wald's entropy is equal to a quarter of the horizon area in units of the
  effective gravitational coupling'',
  arXiv:0712.3206 [hep-th]
\bibitem{Dunne:2004nc}
  G.~V.~Dunne,
  ``Heisenberg-Euler effective Lagrangians: Basics and extensions,''
  arXiv:hep-th/0406216
\bibitem{Preparata:1998rz}
  G.~Preparata, R.~Ruffini and S.~S.~Xue,
  %``The dyadosphere of black holes and gamma-ray bursts,''
  Astron.\ Astrophys.\  {\bf 338}, L87 (1998)
  [arXiv:astro-ph/9810182]
\bibitem{Ruffini:1998df}
  R.~Ruffini,
  ``On the dyadosphere of black holes,''
  [arXiv:astro-ph/9811232]
\bibitem{Page:2006cm}
  D.~N.~Page,
  %``Evidence Against Astrophysical Dyadospheres,''
  Astrophys.\ J.\  {\bf 653}, 1400 (2006)
  [arXiv:astro-ph/0610340]\\
%\bibitem{Page:2006zk}
  D.~N.~Page,
  %``Dyadospheres Don't Develop,''
  [arXiv:astro-ph/0605434]\\
  %%CITATION = ASTRO-PH/0605434;%%
%\bibitem{Page:2006zi}
  D.~N.~Page,
  %``No Astrophysical Dyadospheres,''
  [arXiv:astro-ph/0605432]
\end{thebibliography}
    \end{document}